\documentclass[a4paper]{amsart}
\pdfoutput=1
\usepackage[tight,footnotesize]{subfigure}

\usepackage[numbers,sort]{natbib}

\usepackage{color,graphicx}

\usepackage{units}

\usepackage{overpic}

\usepackage{booktabs}

\definecolor{midnightblue}{rgb}{0.098039, 0.098039, 0.43922}
\definecolor{slateblue}{rgb}{0.41569,0.35294,0.80392}
\definecolor{darkolive}{rgb}{0.333333, 0.419608, 0.184314}
\definecolor{lightblue}{rgb}{0.666667,0.666667,1}
\definecolor{royalblue}{rgb}{0.254902, 0.411765, 0.882353}

\usepackage{mathtools}
\usepackage{amsfonts}

\newcommand{\IOmega}{{\Omega^{-}}}
\newcommand{\EOmega}{{\Omega^{+}}}
\newcommand{\ID}{{\gamma_D^{-}}}
\newcommand{\ED}{{\gamma_D^{+}}}
\newcommand{\Dirichlet}{{\gamma_D}}
\newcommand{\IN}{{\gamma_N^{-}}}
\newcommand{\EN}{{\gamma_N^{+}}} 
\newcommand{\mat}[1]{#1}
\newcommand{\Vek}[1]{\mathbf{#1}}
\newcommand{\DLP}{\Vek{K}}
\newcommand{\FS}{U}
\DeclareMathOperator{\Curl}{\mathbf{curl}}
\DeclareMathOperator{\Div}{div}
\DeclareMathOperator{\Grad}{\mathbf{grad}}
\DeclareMathOperator{\sdiv}{\Div_\Gamma}
\DeclareMathOperator{\diag}{diag}
\DeclareMathOperator{\curl}{curl}
\DeclareMathOperator{\scurl}{\curl_\Gamma}
\DeclareMathOperator{\dif}{d}
\DeclareMathOperator{\rank}{rank}
\DeclareMathOperator{\meas}{vol}
\providecommand{\D}[1]{\dif\!#1}

\begin{document}
%
\title{Simulation of electrical machines -- A FEM-BEM coupling scheme}
\author[L. Kielhorn, T. R{\"u}berg, J. Zechner]{Lars~Kielhorn, Thomas~R{\"u}berg, J{\"u}rgen Zechner}
\address{TailSiT GmbH, Nikolaiplatz 4, 8020 Graz}
\email{info@tailsit.com}
\urladdr{http://tailsit.com}

\begin{abstract} Electrical machines
    commonly consist of moving and stationary parts.  The field simulation of
    such devices can be very demanding if the underlying numerical scheme is
    solely based on a domain discretization, such as in case of the Finite
    Element Method (FEM).  Here, a coupling scheme based on FEM together with
    Boundary Element Methods (BEM) is presented that neither hinges on
    re-meshing techniques nor deals with a special treatment of sliding
    interfaces. While the numerics are certainly more involved the reward is
    obvious: The modeling costs decrease and the application engineer is
    provided with an easy-to-use, versatile, and accurate simulation tool.
  \end{abstract}


\maketitle

\section{Introduction}
Today's development cycles of electric machines, magnetic sensors, or
transformers are intimately connected with numerical simulation. A
cost-effective development and optimization of these devices is hardly viable
without virtual prototyping. The fundamentals of electromagnetic simulations
are the Maxwell's equations and one of the most popular and most versatile
numerical discretization schemes is the Finite Element Method (FEM). While
originally applied to problems in structural mechanics the FEM succeeded also
for electromagnetic problems for more than 30 years.

However, the simplicity of the FEM does not come for free. Since
electric and magnetic fields extend into the unbounded exterior air region one
typically introduces homogeneous boundary conditions some distance away from
the solid parts. Then by expanding the Finite Element grid to parts of the air
region an approximation for the unknown fields can be obtained. Thanks to the
decay properties of the electromagnetic fields this approach is widely
applied, accepted, and justified for many applications. Nevertheless, some
problems remain:
\begin{itemize}
\item Non-physical boundary conditions are imposed on the domain's
  (fictitious) boundary and the introduced modeling error leads to
  contaminated solutions. This might become critical when highly accurate
  simulation results are needed.
\item The meshing of the air region requires a considerable amount of time and
  effort. In many situations the number of elements in the air region even
  exceeds the number of elements used for the solid parts.
\item Electrical devices often contain moving parts. For instance, the
  variation of the rotor/stator positions of an electric motor requires either
  a re-meshing of the air gap or a fundamental modification of the Finite
  Element scheme.
\item The accurate computation of electro-mechanical forces with a FEM-only
  discretization remains a challenge.
\end{itemize}
In this work we are going to present the implementation of a FEM-BEM coupling
scheme in which the air region is handled by the Boundary Element Method (BEM)
while only the solid parts are discretized by the FEM. The BEM is a convenient
tool to tackle unbounded exterior domains and it is well-designed to eliminate
the above mentioned problems. The method is based on the discretization of
Boundary Integral Equations (BIEs) \cite{hsiaoWendland:08} that are defined on
the surface of the computational domain. Hence, no meshing is required for the
air region. Further, the BIEs fulfill the decay and radiation conditions of
the electromagnetic fields such that no additional modeling errors occur. And
finally, a simple, automatic treatment of moving parts is intrinsic to the
presented scheme. The force computation is beyond the scope of this paper but
it is important to note that the BEM computations provide us with additional
information such that these calculations can be carried out with unparalleled
accuracy.

This work is organized as follows: Section \ref{sec:mathematics} recalls the
main idea of the FEM-BEM coupling. The coupling relies mainly on the
theoretical work presented in \cite{hiptmair:02}. For proper definitions of
the mathematical spaces, their traces, and the occurring surface differential
operators we refer to that publication and to the references cited therein. In
section \ref{sec:numericalExamples} the method itself and the presented
preconditioning strategy are verified.  Additionally, applications to
industrial models are presented, where we remark on the incorporation of
periodic constraints. Finally, this work closes with a conclusion in section
\ref{sec:conclusion}.

\section{Symmetric FEM-BEM Coupling\label{sec:mathematics}}
We summarize the FEM-BEM coupling scheme as it has been proposed in
\cite{hiptmair:02}. The governing equations for magnetostatics are
\begin{equation}
  \label{eq:govEqn}
  \Div \Vek{B} = 0 \,, \quad \Curl \Vek{H} = \Vek{j}\,, \quad \Vek{B} = \mu \left(
    \Vek{H} + \Vek{M} \right) \;.
\end{equation}
The first equation is Gauss's law for magnetism, the second equation is
Amp{\'e}res law in differential form, and the third equation is the material
law. It connects the magnetic flux $\Vek{B}$ with the magnetic field $\Vek{H}$
via the -- possibly non-linear -- magnetic permeability $\mu$. The prescribed
quantities are the solenoidal excitation current, $\Vek{j}$, and a given
magnetization $\Vek{M}$. By introducing the non-gauged vector potential
$\Vek{A}$ such that $\Curl \Vek{A} \coloneqq \Vek{B}$ the following
variational formulation holds: 

\emph{Find $\Vek{A} \in H(\Curl,\IOmega)$ such
  that}
\begin{multline}
  \label{eq:varGovEqn}
  \langle \mu^{-1} \Curl \Vek{A}, \Curl \Vek{A}^\prime \rangle_\IOmega -
  \langle \mu^{-1} \IN \Vek{A}, \ID \Vek{A}^\prime \rangle_\Gamma = 
  \\ \langle \Vek{j}, \Vek{A}^\prime \rangle_\IOmega + \langle \Vek{M},\Curl
  \Vek{A}^\prime \rangle_\IOmega
\end{multline}
\emph{for all test-functions $\Vek{A}^\prime \in H(\Curl,\IOmega)$.}  

Above, we have used the notation \mbox{$\langle \Vek{u},\Vek{w} \rangle_M
  \coloneqq \int_M \Vek{u}\cdot\Vek{w} \, \D{M}$}. The interior domain is
$\IOmega \subset \mathbb{R}^3$ and $\Gamma \coloneqq \partial \IOmega$ is its
boundary. The complementary unbounded exterior air region is given by $\EOmega
= \mathbb{R}^3 \setminus \overline{\IOmega}$. The space $H(\Curl ,\IOmega)$
denotes the space of square integrable vector functions with a square
integrable curl \cite{monk2003finite}. Further, $\gamma$ denotes the trace
operator. The Dirichlet and Neumann traces for the interior as well as for the
exterior domain are given by
\begin{equation}
  \label{eq:traceOperators}
  \begin{aligned}
  \gamma^\pm_D \Vek{A} &\coloneqq \lim_{\Omega^\pm \ni \tilde{\Vek{x}} \to
    \Vek{x} \in \Gamma} \Vek{n}(\Vek{x}) \times (\Vek{A}(\tilde{\Vek{x}}) \times
  \Vek{n}(\Vek{x}))    \\
  \gamma^\pm_N \Vek{A} &\coloneqq \lim_{\Omega^\pm \ni \tilde{\Vek{x}} \to
    \Vek{x} \in \Gamma} \Curl_{\tilde{\Vek{x}}} \Vek{A}(\tilde{\Vek{x}})
  \times \Vek{n}(\Vek{x})
  \end{aligned}
\end{equation}
with $\Vek{n}$ being the normal vector outward to $\IOmega$. Schemes that rely
solely on Finite Element Methods commonly neglect the boundary term in
\eqref{eq:varGovEqn}, i.e., either $\IN \Vek{A}$ or $\ID \Vek{A}^\prime$ are
ignored such that the variational formulation corresponds to a homogeneous
Neumann- or Dirichlet-problem, respectively.  However, in a coupled
formulation the interior traces are expressed by their exterior counterparts
via appropriate transmission conditions. For the given physical model, these
transmission conditions read
\begin{equation}
  \label{eq:transmission}
  \ID \Vek{A} = \ED \Vek{A} \qquad \mu^{-1} \IN \Vek{A} = \mu_0^{-1} \EN \Vek{A} 
\end{equation}
with $\mu_0$ denoting the vacuum permeability. The vector potential $\Vek{A}$
fulfills the following representation formula
\cite[Ch. 6.2]{coltonKress2012inverse}
\begin{equation}
  \label{eq:strattonChu}
  \begin{aligned}
  \Vek{A}(\tilde{\Vek{x}}) = 
  &
  \int_\Gamma \FS (\tilde{\Vek{x}}-\Vek{y}) \, \EN \Vek{A}(\Vek{y}) \,
  \D{s_{\Vek{y}}}\\
  &-\Curl_{\tilde{\Vek{x}}} \int_\Gamma
  \FS (\tilde{\Vek{x}}-\Vek{y}) \, (\Vek{n} \times \Vek{A})(\Vek{y}) \, \D{s_{\Vek{y}}} \\
  &+\Grad_{\tilde{\Vek{x}}} S(\Vek{A})
\end{aligned}
\end{equation}
for all points $\tilde{\Vek{x}} \in \Omega^+$, $\Vek{y} \in \Gamma$, and with
$\FS (\tilde{\Vek{x}}-\Vek{y}) \coloneqq \nicefrac{1}{4 \pi
  |\tilde{\Vek{x}}-\Vek{y}|}$ being the fundamental solution for the
Laplace operator. The last term in \eqref{eq:strattonChu} is 
\begin{equation}
  \label{eq:gaugeBIE}
  S(\Vek{A}) \coloneqq \int_\Gamma \FS (\tilde{\Vek{x}}- \Vek{y}) \, (\Vek{n} \cdot
  \Vek{A})(\Vek{y}) \, \D{s_{\Vek{y}}} \;.
\end{equation}
Applying the traces $\ED$, and $\EN$ to \eqref{eq:strattonChu} we arrive at
the set of weak boundary integral equations
\begin{equation}
  \label{eq:bie}
  \begin{aligned}
    \langle \gamma_D^+ \Vek{A}, \boldsymbol{\zeta} \rangle_\Gamma &=
    \langle \DLP(\gamma_D^+\Vek{A}),\boldsymbol{\zeta} \rangle_\Gamma - \langle \Vek{V}(\gamma_N^+ \Vek{A}), \boldsymbol{\zeta} \rangle_\Gamma 
    \\
    \langle \gamma_N^+ \Vek{A}, \Vek{w} \rangle_\Gamma &=
    \langle \Vek{N}(\gamma_D^+ \Vek{A}), \Vek{w} \rangle_\Gamma - \langle
    \DLP^*(\gamma_N^+ \Vek{A}), \Vek{w} \rangle_\Gamma
    \;.
  \end{aligned}
\end{equation}
Above, $\Vek{V}$ denotes the Maxwell single layer potential, $\Vek{N}$ is the
hypersingular operator, and $\DLP$, $\DLP^*$ are the Maxwell double layer
potential and its adjoint, respectively
\begin{equation}
  \label{eq:BIEoperators}
  \begin{aligned}
    \Vek{V} (\EN \Vek{A}) &\coloneqq \gamma_D^+
    \boldsymbol{\psi}_{\Vek{A}}(\EN \Vek{A}) &\qquad \Vek{K} (\ED \Vek{A})
    &\coloneqq \gamma_D^+ \boldsymbol{\psi}_{\Vek{M}}
    (\ED \Vek{A}) \\
    \Vek{K}^* (\EN \Vek{A}) &\coloneqq \gamma_N^+ \boldsymbol{\psi}_{\Vek{A}}
    (\EN \Vek{A}) &\qquad \Vek{N} (\ED \Vek{A}) &\coloneqq \gamma_N^+
    \boldsymbol{\psi}_{\Vek{M}}(\ED \Vek{A}) \;.
  \end{aligned}
\end{equation}
The potentials $\boldsymbol{\psi}_{\Vek{A}}$ and $\boldsymbol{\psi}_{\Vek{M}}$
read 
\begin{equation}
  \label{eq:psiOp}
  \begin{aligned}
  (\boldsymbol{\psi}_{\Vek{A}} \Vek{u})(\tilde{\Vek{x}}) &\coloneqq \int_\Gamma
  \FS(\tilde{\Vek{x}}-\Vek{y}) \, \Vek{u}(\Vek{y}) \, \D{s_{\Vek{y}}} \\
  (\boldsymbol{\psi}_{\Vek{M}} \Vek{v})(\tilde{\Vek{x}}) &\coloneqq
  \Curl_{\tilde{\Vek{x}}} \boldsymbol{\psi}_{\Vek{A}}(\Vek{n} \times
  \Vek{v})(\tilde{\Vek{x}})
  \end{aligned}
\end{equation}
for all points $\tilde{\Vek{x}} \notin \Gamma$. For further informations on the
above introduced boundary integral operators we refer to
\cite{hiptmair:02}. Eqn. \eqref{eq:bie} holds for
$\boldsymbol{\zeta} \in H(\sdiv 0,\Gamma) \subset H(\sdiv,\Gamma)$ and
$\Vek{w} \in H(\scurl,\Gamma)$. While $\Vek{w}$ is an element of the trace
space of $H(\Curl,\IOmega)$, the function $\boldsymbol{\zeta}$ is taken from
the space of vector fields with vanishing surface divergence, i.e.,
$\sdiv \boldsymbol{\zeta} \equiv 0$. The use of this special space is
motivated by the fact that the potential in \eqref{eq:gaugeBIE} has no
physical meaning within this physical context and, therefore, shall be removed
from the final numerical scheme. Because of $\Curl \Grad S \equiv 0$ the
potential drops out naturally when the Neumann trace is applied to
\eqref{eq:strattonChu}. Contrary, for the Dirichlet trace it vanishes only in
the weak setting. Integration by parts then gives
\begin{equation}
  \label{eq:intbyparts}
  \langle \gamma_D^+ \Grad_{\tilde{\Vek{x}}} S(\Vek{A}), \boldsymbol{\zeta}
  \rangle_\Gamma = \langle \Grad_\Gamma S(\Vek{A}),\boldsymbol{\zeta}
  \rangle_\Gamma = -\langle S(\Vek{A}),\sdiv \boldsymbol{\zeta}
  \rangle_\Gamma
  = 0 \;.
\end{equation}
Due to $\sdiv \Curl_\Gamma \varphi \equiv 0$ the divergence-free constraint
can be imposed via the surface rotation $\Curl_\Gamma$ of some continuous
scalar function $\varphi$. Let $\boldsymbol{\lambda} \coloneqq \EN \Vek{A}$
abbreviate the exterior Neumann trace. We then define the vector field
\begin{equation}
  \label{eq:ansatzLambda}
  \boldsymbol{\lambda} \coloneqq \Curl_\Gamma \varphi \,, \quad \varphi \in
  H^{1/2}(\Gamma) \;.
\end{equation}

Using the transmission conditions \eqref{eq:transmission}, the variational
formulations \eqref{eq:varGovEqn} and \eqref{eq:bie} are combined. Introducing
the relative permeability $\mu_r \coloneqq \nicefrac{\mu}{\mu_0}$ together
with $\ID = \ED \eqqcolon \Dirichlet$, the coupled variational form reads:
\emph{Find $\Vek{A} \in H(\Curl,\IOmega)$, $\varphi \in H^{1/2}(\Gamma)$ such
  that}
\begin{equation}
  \label{eq:coupledVarForm}
  \begin{aligned}
    \langle \mu^{-1}_r \Curl \Vek{A}, \Curl \Vek{A}^\prime \rangle_\IOmega -
    \langle \Vek{N}(\Dirichlet \Vek{A}), \Dirichlet \Vek{A}^\prime
    \rangle_\Gamma +&\langle \DLP^* \Curl_\Gamma \varphi, \Dirichlet
    \Vek{A}^\prime \rangle_\Gamma\\ 
    = \mu_0 \langle \Vek{j}&,\Vek{A}^\prime \rangle_\IOmega + \mu_0 \langle
    \Vek{M},\Curl \Vek{A}^\prime \rangle_\IOmega
    \\
    \langle \left( \DLP - \Vek{Id} \right) (\Dirichlet \Vek{A}),\Curl_\Gamma
    \psi \rangle_\Gamma - \langle \Vek{V} \Curl_\Gamma \varphi&, \Curl_\Gamma
    \psi \rangle_\Gamma = 0
  \end{aligned}
\end{equation}
\emph{for all functions $\Vek{A}^\prime \in H(\Curl,\IOmega)$, $\psi \in
  H^{1/2}(\Gamma)$.} Note that the variational form \eqref{eq:coupledVarForm}
is symmetric because of
\begin{equation}
  \label{eq:symmetryVarForm}
  \langle \DLP^* \boldsymbol{\lambda},\Vek{w} \rangle_\Gamma =
  \langle (\DLP-\Vek{Id}) \Vek{w},\boldsymbol{\lambda} \rangle_\Gamma 
\end{equation}
for all $\Vek{w} \in H(\scurl,\Gamma)$, $\boldsymbol{\lambda} \in H(\sdiv
0,\Gamma)$ \cite{hiptmair:02}.

Next, a geometrically conformal triangulation $\overline{\Omega_h} \coloneqq
\cup_{n=1}^N K_n \approx \IOmega$ is introduced. The $N$ mesh elements $K_n$
may consist of tetrahedra, hexahedra, prisms, and pyramids. Additionally, this
triangulation induces a surface mesh $\Gamma_h$ of $\Gamma$ composed of
triangles and/or quadrilaterals. N{\'e}d{\'e}lec elements
\cite{nedelec:80,bergot2013high} are used for a conformal discretization of
$H(\Curl,\IOmega)$. The linear system corresponding to
\eqref{eq:coupledVarForm} then reads
\begin{equation}
  \label{eq:linSystemGeneral}
  S \underline{x} = \underline{g}
\end{equation}
with
\begin{equation}
  \label{eq:linSystem}
    S = \begin{bmatrix}
      A + R^\top N R & R^\top K^\top G \\
      G^\top K R & -G^\top V G
    \end{bmatrix} , \,
    \underline{x} =
    \begin{bmatrix}
      \underline{a} \\ \underline{\varphi}
    \end{bmatrix} , \,
    \underline{g} = 
    \begin{bmatrix}
      \underline{f} \\ 0
    \end{bmatrix} \,.
\end{equation}
Above, $A$ is the matrix corresponding to the discretization of the
$\Curl$-operator (see Eqn. \eqref{eq:varGovEqn}), and $N$, $K$, and $V$ are
matrix representations of the formerly introduced BEM-operators. The given
right hand side is $\underline{f} = \underline{f}(\Vek{j};\Vek{M})$ and it
represents the right hand side terms of the first equation in
\eqref{eq:coupledVarForm}. Fast Boundary Element Methods are utilized in order
to obtain \emph{data sparse} representations of the originally dense BEM
matrices \cite{hackbusch:09,greengardRokhlin:97}. Their use is mandatory since
they reduce the quadratic complexity $O(N^2)$ of the Boundary Element Method
to quasi-linear costs $O(N \log N)$ or even to linear costs $O(N)$, respectively.  It
remains to comment on the matrices $R$ and $G$, respectively. In
\eqref{eq:linSystem}, $R$ is a restriction matrix that extracts the boundary
coefficients. The matrix $G$ maps the surface curls of piecewise linear
continuous functions to elements of $H(\sdiv,\Gamma)$. It is sometimes denoted
as \emph{topological gradient} \cite[Ch. 3]{hiptmair:02FEM}. The entries of
$G$ are given by $G[\ell,k] = \pi_\ell( \Curl_\Gamma \varphi_k )$. Here,
$\pi_\ell$ denotes the functional that defines the degrees of freedom for the
Raviart-Thomas element which is used for the discretization of
$H(\sdiv,\Gamma)$ \cite{raviart:1977}. For the lowest order space this
functional is $\pi_\ell(\Vek{v}) \coloneqq \int_{E_\ell} \left( \Vek{n} \times
  \Vek{v} \right) \cdot \D{\Vek{s}}$ with $E_\ell$ being the $\ell$-th edge of
the surface mesh $\Gamma_h$.

\begin{figure}[h]
  \centering
  \begin{overpic}[width=0.5\textwidth]{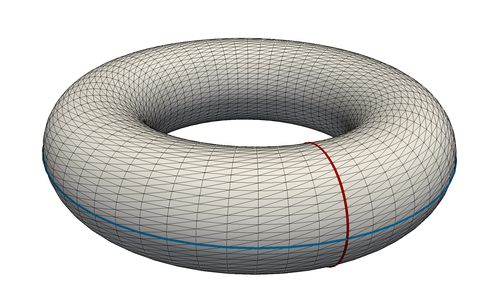}
    \footnotesize
    \put(69,25){$\gamma_1^\prime$}
    \put(50,14){$\gamma_1$}
  \end{overpic}
  \caption{Torus with its generating cycles $\gamma_1$ and $\gamma_1^\prime$}
  \label{fig:cohomology}
\end{figure}

The construction of the numerical scheme is almost complete. However, one
challenge remains: In most applications, the interior domain $\IOmega$ is
non-simply connected. The torus in Fig. \ref{fig:cohomology} depicts such a
domain since the two loops $\gamma_1^\prime$ and $\gamma_1$ cannot be
contracted on the surface to a point. The definition \eqref{eq:ansatzLambda}
introduces a scalar potential for the exterior domain such that we have
$\Vek{H}^+ = -\Grad \varphi$ for the exterior magnetic field. Let $\Sigma$ be
some oriented cross section area and $\gamma=\partial \Sigma$ a closed
path. The integral form of Amp{\'e}re's law is
\begin{equation}
  \label{eq:ampere}
  \oint_{\gamma} \Vek{H}^+ \cdot \D{\boldsymbol{\ell}} = \int_\Sigma \Vek{j}
  \cdot \D{\boldsymbol{\Sigma}} = I
\end{equation}
with the total current density $I$. We assume that the torus in
Fig. \ref{fig:cohomology} features a total current density $I \neq
0$. Obviously Amp{\'e}re's law is violated if the path $\gamma^\prime_1$ is
chosen. In this case the path integral of the gradient field evaluates to
zero. This contradicts \eqref{eq:ampere} and the choice
\eqref{eq:ansatzLambda} has to be augmented
\begin{equation}
  \label{eq:augmentedLambda}
  \boldsymbol{\lambda} \coloneqq \Curl_\Gamma \varphi + \sum_{m=1}^M \alpha_m
  \boldsymbol{\eta}_m \;.
\end{equation}
In \eqref{eq:augmentedLambda}, the functions $\boldsymbol{\eta}_m \in H(\sdiv
0,\Gamma)$ are \emph{current sheets} along paths $\gamma_m$. They feature the
jump 
\begin{equation}
  \label{eq:jump}
  \oint_{\gamma_m^\prime} (\Vek{n} \times \boldsymbol{\eta}_m) \cdot
  \D{\boldsymbol{\ell}} = 1
\end{equation}
across the path $\gamma_m$. The number $M$ of relevant paths is given by the
number of holes drilled through $\IOmega$. An algorithm for the construction
of these paths can be found in \cite{hiptmair2002generators}. Inserting
\eqref{eq:augmentedLambda} into the variational form \eqref{eq:coupledVarForm}
results in the modified system
\begin{equation}
  \label{eq:linSystemCohom}
  \begin{bmatrix}
    S & F_{\boldsymbol{\eta}} \\
    F^\top_{\boldsymbol{\eta}} & H_{\boldsymbol{\eta}}
  \end{bmatrix} \cdot 
  \begin{bmatrix}
    \underline{x} \\ \underline{\alpha}
  \end{bmatrix} =
  \begin{bmatrix}
    \underline{g} \\ 0
  \end{bmatrix}  \;.
\end{equation}
With $S_{\boldsymbol{\eta}} \coloneqq F_{\boldsymbol{\eta}}
H_{\boldsymbol{\eta}}^{-1} F^\top_{\boldsymbol{\eta}}$,
\eqref{eq:linSystemCohom} can alternatively be written as
\begin{equation}
  \label{eq:linSystemSchur}
  \widehat{S} \underline{x} = \underline{g} \,, \quad \widehat{S}
  \coloneqq S - S_{\boldsymbol{\eta}} \;.
\end{equation}
The matrix $S_{\boldsymbol{\eta}}$ is a $\rank(M)$ perturbation of
the original system \eqref{eq:linSystemGeneral}. Since $M \ll \dim(S)$, this
perturbation does not significantly alter the spectral properties of
$S$. Hence, the upcoming preconditioning strategy neglects the influence of
$S_{\boldsymbol{\eta}}$. 

\begin{figure}[htbp]
  \centering
  \includegraphics[width=0.6\textwidth]{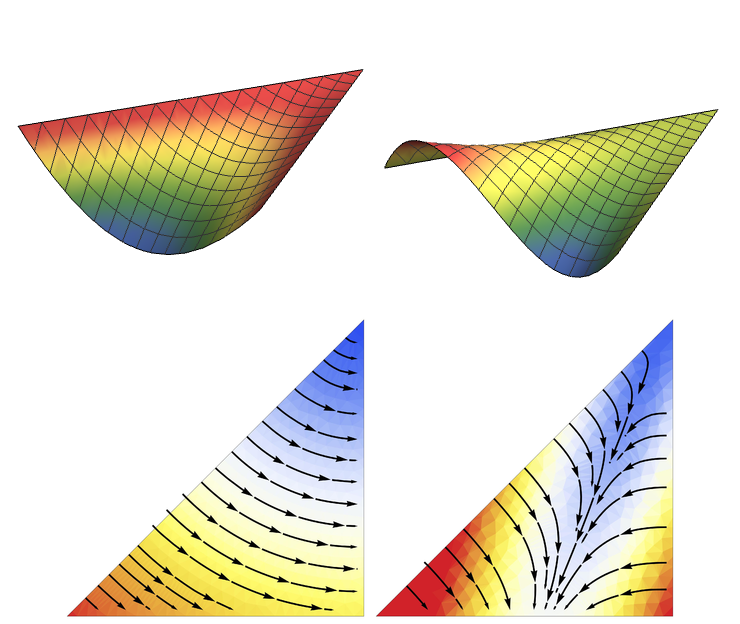}
  \caption{Hierarchical functions on the triangle. \emph{Upper row:}
    $2^{\text{nd}}$ and $3^{\text{rd}}$ order functions for
    $H^{1/2}(\Gamma)$. \emph{Lower row:} $1^{\text{st}}$ and $2^{\text{nd}}$
    order functions for $H(\scurl,\Gamma)$.}
  \label{fig:hierarchicalFunctions}
\end{figure}

The linear system \eqref{eq:linSystemGeneral} is symmetric but not positive
definite. We therefore apply a MINRES solver \cite{minres1975} to the
preconditioned system. The block preconditioner is given by $P_S^{-1} =
\diag(P^{-1}_{\text{AMS}},P^{-1}_V)$ where $P^{-1}_{\text{AMS}}$ is an AMG/AMS
preconditioner. This preconditioner is based on algebraic multigrid methods
(AMG) as they have been developed for the discretization of Finite Element
spaces like, e.g., $H^1(\Omega)$. However, since we are dealing with
N{\'e}d{\'e}lec-type Finite Element spaces the standard AMG preconditioners
cannot be applied directly. An enhancement based on auxiliary space methods
has been developed in \cite{hiptmairXu:07}. It is denoted as \emph{auxiliary
  Maxwell space} preconditioner (AMS) and appears to be the natural choice for
magnetostatic and eddy-current problems. The BEM preconditioner $P^{-1}_V$ is
based on operator preconditioning techniques \cite{of:06}. Due to the use of
hierarchical function spaces \cite{bergot2013high}
(cf. Fig.~\ref{fig:hierarchicalFunctions}), subspace correction methods are
used in case of higher order schemes \cite{xu1992iterative}.

\section{Numerical Examples\label{sec:numericalExamples}}
\subsection{Verification} The numerical scheme is verified by means of an
academic example. We consider a magnetized unit sphere, i.e., $\IOmega
\coloneqq \{ \Vek{x} \colon |\Vek{x}| < 1 \}$. The sphere has a constant
magnetization $\Vek{M} = M_0 \Vek{e}_3$ in $x_3$-direction with $M_0 = 1$.

\begin{figure}[htbp]
  \centering
  \includegraphics[width=0.8\textwidth]{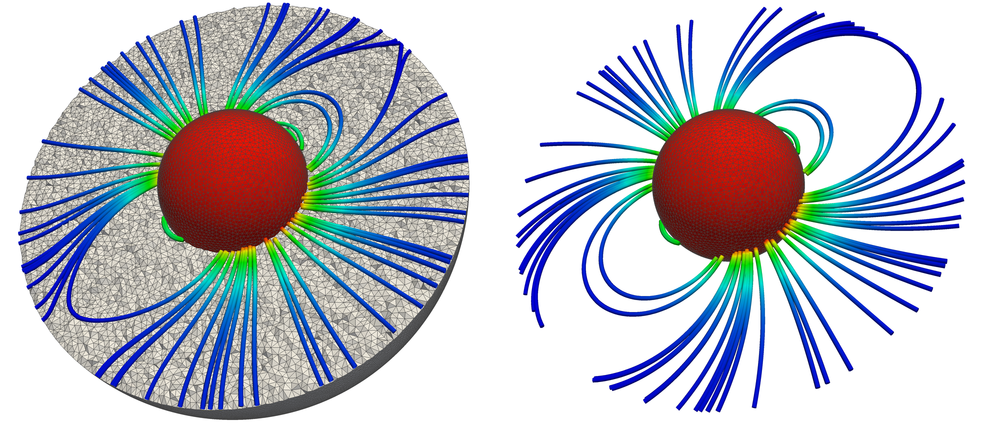}
  \caption{Comparison between FEM and FEM-BEM. \emph{Left:} FEM solution
    $\Vek{B}$ with field lines artificially normal to the
    boundary. \emph{Right:} FEM-BEM solution shows correct behavior.}
  \label{fig:comparisonFEMvsFEMBEM}
\end{figure}

The analytic solution for the magnetic flux is $\Vek{B}^- = \nicefrac{2
  \mu_0}{3} \Vek{M}$ and $\varphi(\Vek{x}) = \nicefrac{x_3}{3}$ for the scalar
potential \cite[Ch. 5.10]{jackson99}.

\begin{figure}[htbp]
  \centering
  \includegraphics{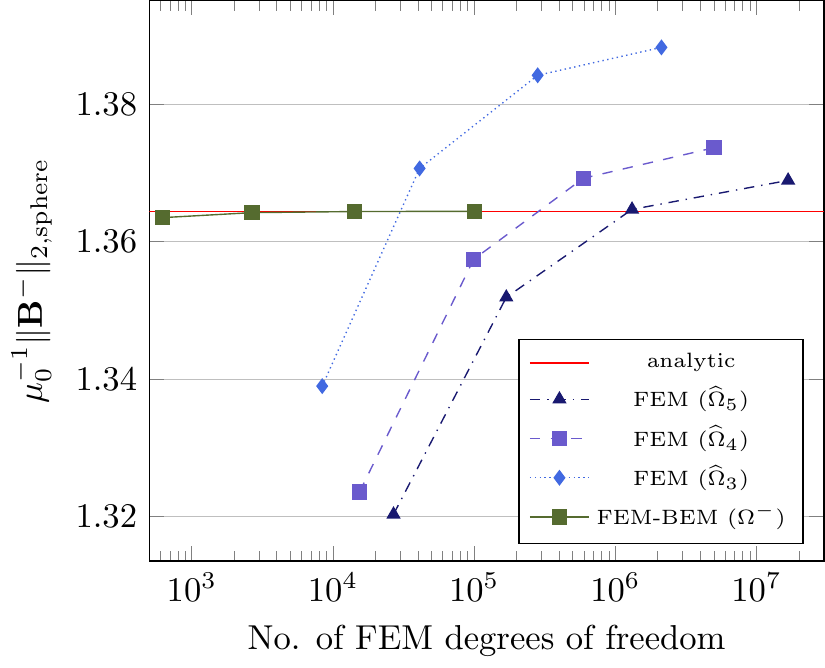}
  \caption{FEM-BEM on $\IOmega$ vs. FEM on $\widehat{\Omega}_R$}
  \label{fig:magnetizedSphereLowestOrder}
\end{figure}

In Fig.~\ref{fig:magnetizedSphereLowestOrder} the $L_2$-norms for the computed
magnetic fluxes are compared against the analytic solution $\mu_0^{-1} \|
\Vek{B}^- \|_{2,\text{sphere}} = |\Vek{B}^-| \sqrt{\meas(\IOmega)} \approx
1.3644$. In addition, pure FEM computations have been taken out for various
fictitious domains $\widehat{\Omega}_R \coloneqq \IOmega \cup \{ \Vek{x}
\colon 1 \le |\Vek{x}| < R \}$ with homogeneous Neumann boundary conditions on
$\partial \widehat{\Omega}_R$. Curved tetrahedral elements have been used for
the experiments. Clearly, the modeling error prevents the FEM computations
from converging to the correct result (cf.
Fig. \ref{fig:comparisonFEMvsFEMBEM}). Even $20$ millions of FEM degrees of
freedoms (dofs) for the finest grid on $\widehat{\Omega}_5$ are outperformed
by just 600 FEM dofs that are used within the FEM-BEM scheme for the coarsest
grid on $\IOmega$.

However, while the results from Fig.~\ref{fig:magnetizedSphereLowestOrder} are
proving the correctness of the FEM-BEM formulation, the comparison with the
FEM-only computations is slightly unfair. This is due to the nature of the
analytic solution. It is constant for the magnetic flux within the sphere and
linear for the scalar potential on the boundary. Both fields can be
represented exactly by the used FE spaces. Hence, the errors for the FEM-BEM
formulation in Fig.~\ref{fig:magnetizedSphereLowestOrder} are just due to the
geometrical error induced by the triangulation. Therefore, in the following
example the FEM-BEM scheme is not only applied to the magnetized sphere
$\IOmega$ but to the domain $\widehat{\Omega}_4$ that includes some air
region. In this case the magnetic flux in the air region is given by
$\Vek{B}^+ = -\mu_0 M_0 \Grad \tfrac{x_3}{3 |\Vek{x}|^3}$ which is not covered
by the Finite Element spaces.

\begin{figure}[htbp]
  \centering
  \includegraphics{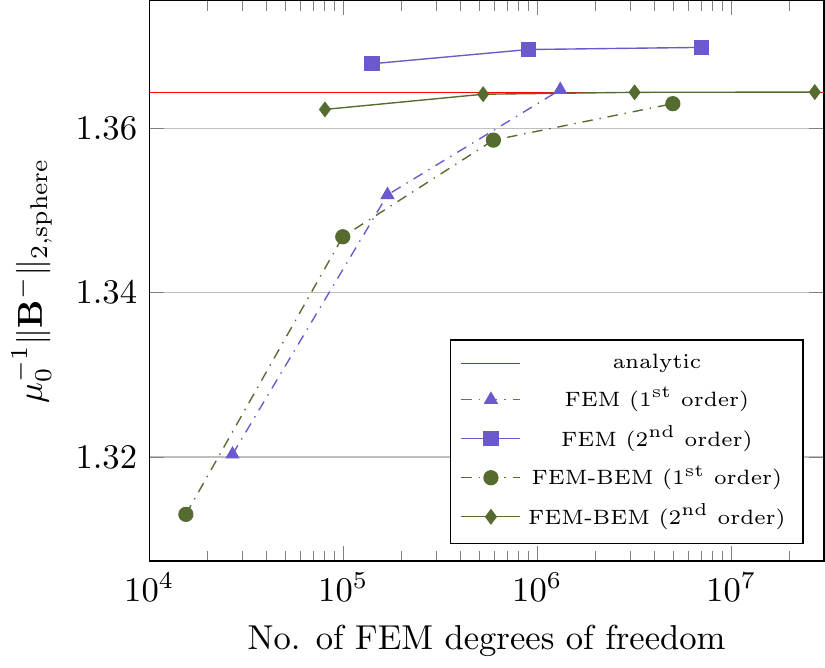}
  \caption{FEM-BEM on $\widehat{\Omega}_4$ vs. FEM on $\widehat{\Omega}_5$}
  \label{fig:magnetizedSphere}
\end{figure}

Fig. \ref{fig:magnetizedSphere} shows the results for the coupled scheme
together with those for the pure FEM computations. Again, we notice that there
is convergence towards the correct solution but now the geometric error is
dominated by the approximation error. Since the lowest order FE spaces exhibit
only linear convergence, higher order schemes are also exploited. Their use is
highly beneficial: Even the results for the second coarsest grid reveal an
accuracy that cannot be achieved by lowest order discretizations on any of the
used grids.

\subsection{Preconditioning} 
\begin{figure}[htbp]
  \centering
  \includegraphics[width=0.95\textwidth]{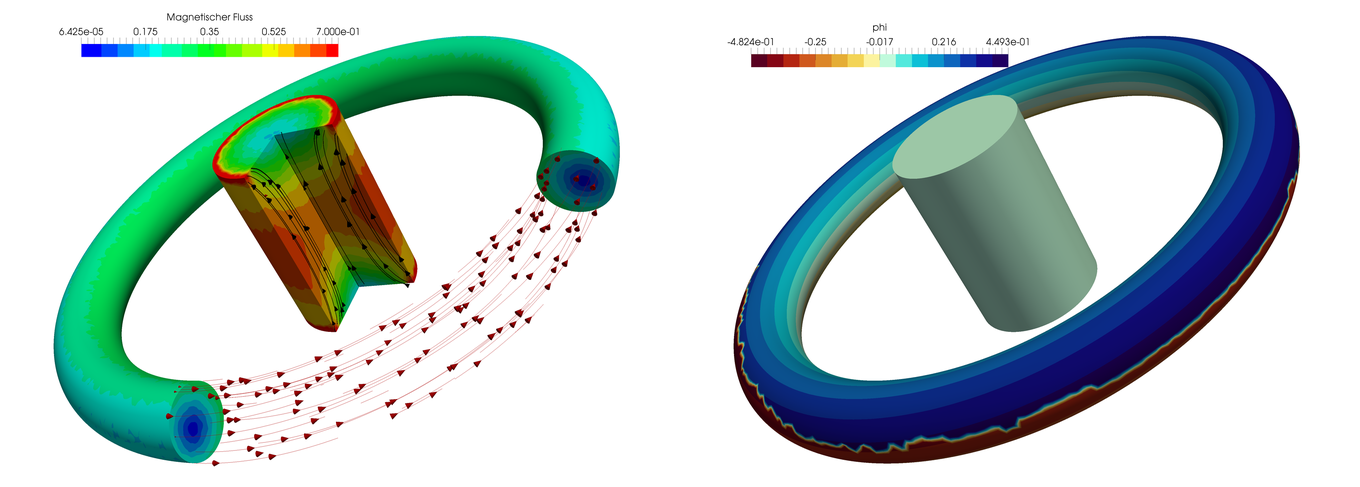}
  \caption{\emph{Left:} Magnetic flux $\Vek{B}$ and excitation current
    $\Vek{j}$. \emph{Right:} Potential $\varphi$ with jump across generating cycle.}
  \label{fig:currentInTorusB}
\end{figure}

Now that the scheme has been verified, the
performance of the preconditioner is investigated. In
Fig. \ref{fig:currentInTorusB}, a circular current loop surrounds a magnetic
core. The model is discretized with three different grids (see
Tab.~\ref{tab:gridConfigs}).

\begin{table}[htbp]
  \centering
  \caption{Discretizations}
  \begin{tabular}{rrrr}
    \toprule
    $L_i$ & Mesh size & $N_{\text{FEM}}$ & $N_{\text{BEM}}$ \\
    \midrule
    $1$ & $0.60$ & $5\,402$   & $706$ \\
    $2$ & $0.39$ & $27\,251$  & $2\,872$ \\
    $3$ & $0.19$ & $166\,748$ & $12\,049$ \\
    \bottomrule
  \end{tabular}
  \label{tab:gridConfigs}
\end{table}
  
In Tab. \ref{tab:precond} the iteration numbers are given for the three
refinement levels $L_i$ as well as for varying permeabilities $\mu_r$. The
relative solver tolerance has been set to $\varepsilon_r =
10^{-8}$. Obviously, the preconditioner is very effective since it reveals
only a slight dependence on the mesh size and on the jumps of material
coefficients.

\begin{table}[htbp]
  \centering
  \caption{Iteration numbers}
  \begin{tabular}{cccc}
    \toprule
    $\mu_r$ & $L_1$ & $L_2$ & $L_3$ \\
    \midrule
    $10^0$ & $47$ & $51$ & $63$ \\
    $10^2$ & $78$ & $88$ & $117$ \\
    $10^5$ & $84$ & $92$ & $119$ \\
    \bottomrule
  \end{tabular}
  \label{tab:precond}
\end{table}

\subsection{Periodicities}
\begin{figure}[htbp]
  \centering
  \includegraphics[width=0.35\textwidth]{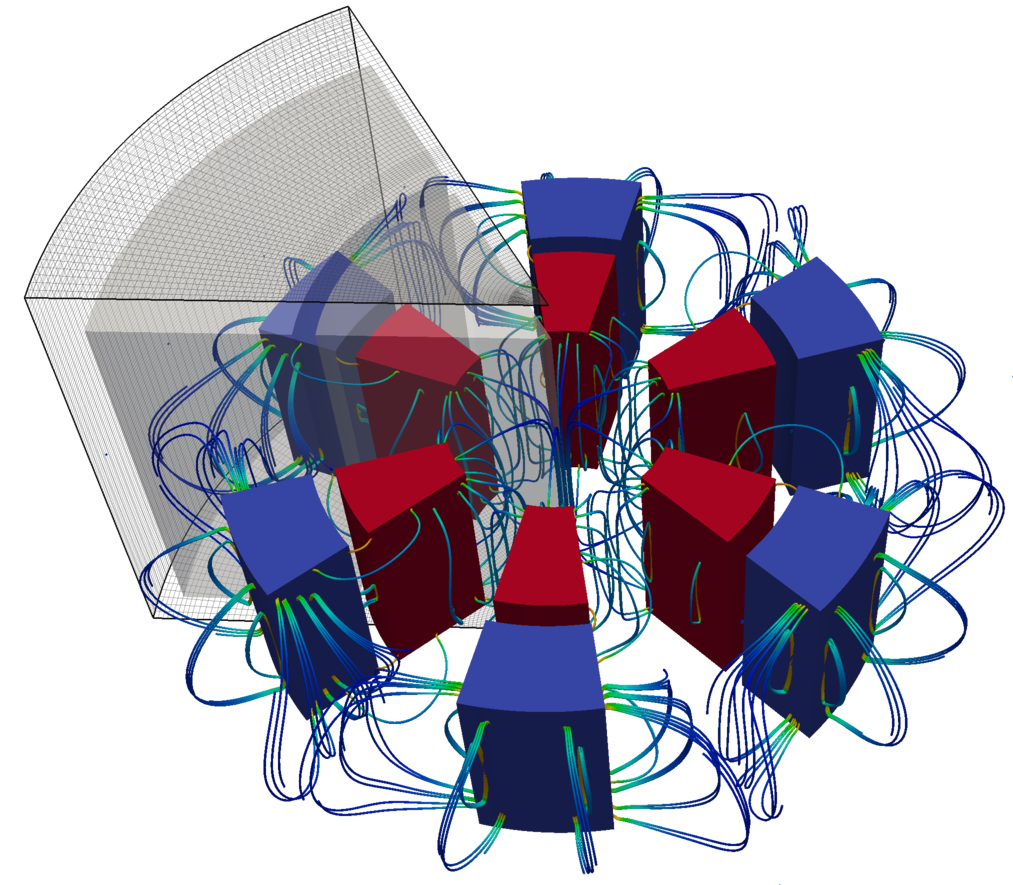}    
  \caption{Device with periodicities. Magnetic flux $\Vek{B}$. $1.5 \cdot
    10^6$ FEM-dofs and $1.4 \cdot 10^4$ BEM-dofs}
  \label{fig:simpleMotor}
\end{figure}

For industrial applications it is important to deal with models that feature
geometrical periodicities. Fig. \ref{fig:simpleMotor} shows a simple motor
model that exploits periodicities. Only a sixth of the model has been
discretized. While it is quite simple to impose periodic constraints into
Finite Element schemes, the situation is a bit more complicated with Boundary
Elements.

\begin{figure}[htbp]
  \centering
  \includegraphics{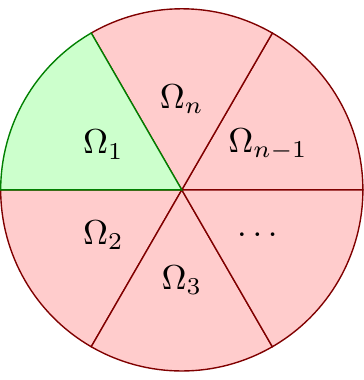}
  \caption{Periodicities}
  \label{fig:periodicities}
\end{figure}

Fig.~\ref{fig:periodicities} sketches a model with periodicities. Due to the
non-local kernel function $\FS (\Vek{x}-\Vek{y})$ interactions on domains
$\Omega_i$, $i = 1,\ldots,n$ with all other domains $\Omega_j$ occur. The
resulting BEM matrix $V$ formally has $n\times n$ blocks $V_{i,j}$ but
the periodicities render $V$ as block circulant matrix
\begin{equation}
  \label{eq:periodicBEM}
  \mat{V} =
  \begin{bmatrix}
    \mat{V}_{1,1}   & \mat{V}_{1,2} & \mat{V}_{1,3} &\dots  &\mat{V}_{1,n} \\
    \mat{V}_{1,n}   & \mat{V}_{1,1} & \mat{V}_{1,2} &\dots  &\mat{V}_{1,n-1} \\
    \mat{V}_{1,n-1} & \mat{V}_{1,n} & \mat{V}_{1,1} &\dots & \mat{V}_{1,n-2} \\ 
    & \ddots & \ddots & \ddots & \\
    \mat{V}_{1,2} & \mat{V}_{1,3} & \mat{V}_{1,4} & \dots & \mat{V}_{1,1}
  \end{bmatrix}\;.
\end{equation}
Hence, only the matrices $V_{1,i}$, $i=1,\ldots,n$ need to be computed and
stored.
If we assume a periodic excitation, we get for the solution vector blocks
$\underline{x}_1 = \ldots = \underline{x}_n \eqqcolon \underline{x}$ and
therefore the matrix-vector product \eqref{eq:periodicBEM} simplifies to
\begin{equation}
  \label{eq:periodicBEMperiodicRhs}
  \underline{y} = \sum_{i=1}^{n} V_{1,i} \, \underline{x} \;.
\end{equation}
The above expression simplifies further if the properties of the
Galerkin-scheme are taken into account. The BEM matrices in
Eqn. \eqref{eq:linSystem} are symmetric and this symmetry reduces the
complexity even further. In case of non-periodic excitations, techniques like
the Fast Fourier Transform are exploited to define a fast matrix-vector
product for the matrix \eqref{eq:periodicBEM}. More information about the
handling of periodicities can be found in
\cite{kurzPeriodicity:03,allgower:98}.

\subsection{Magnetic valve with moving armature}
\begin{figure}[htbp]
  \centering
  \includegraphics[width=0.8\textwidth]{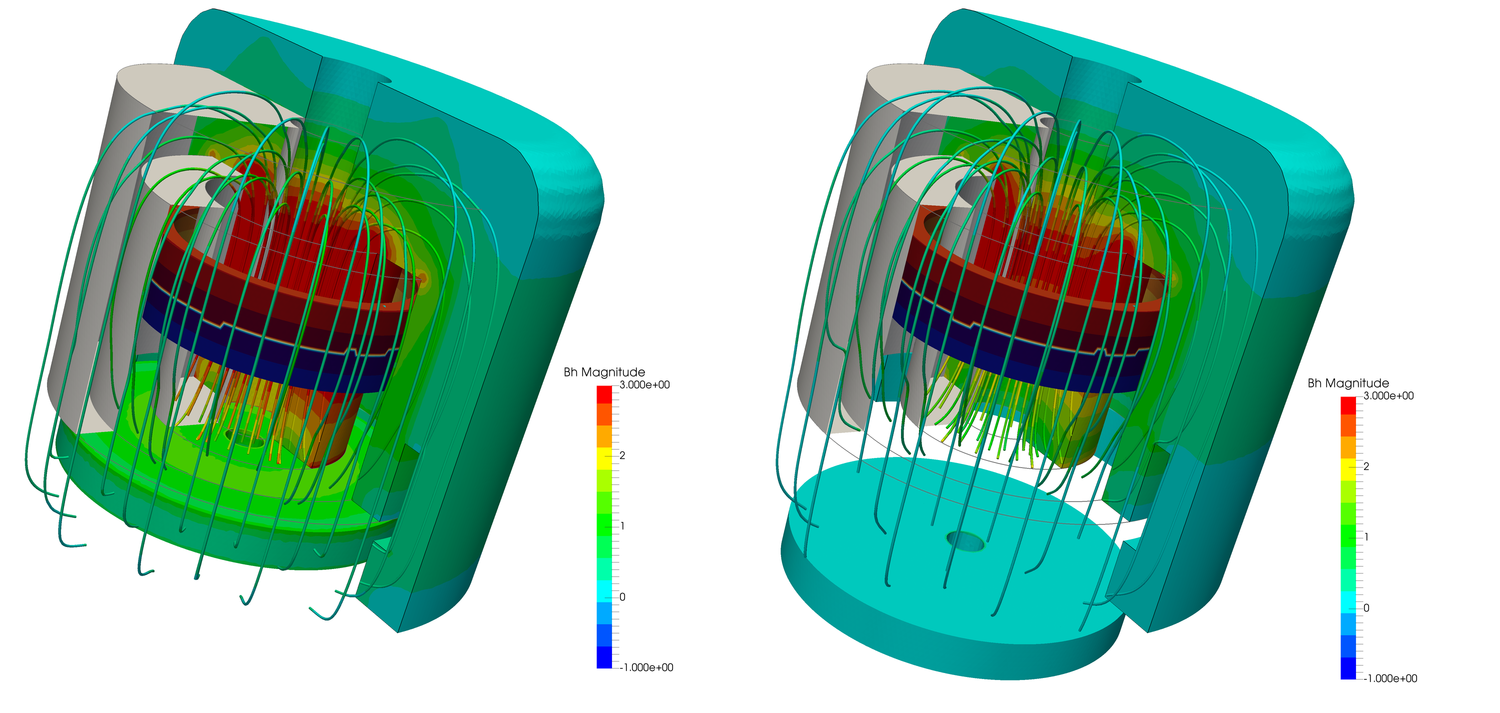}    
  \caption{Moving armature: $\Vek{B}$-field and scalar potential $\varphi$ on
    the toroidal coil. Note the jump of the scalar potential on the coil's
    surface.}
  \label{fig:magneticValve}
\end{figure}

Finally, Fig. \ref{fig:magneticValve} shows a magnetic valve that consists of
approximately $2.4 \cdot 10^6$ degrees of freedom in the FEM domain and of
$1.1\cdot 10^5$ degrees of freedom for the BEM part. At its bottom the valve
has a moving armature and the excitation is given by a circular current
$\Vek{j}$. The relative solver tolerance is set to $\varepsilon_r = 10^{-6}$
and $236$ iterations are required to solve the system for the initial
configuration (Fig. \ref{fig:magneticValve}). Less iterations are then needed
for the moving parts because the previous solution provides a good start
vector.

The magnetic valve, consists of three distinct components: The armature at its
bottom, the toroidal coil, and the remaining composite components like, e.g.,
some permanent magnets and the housing. Thus, for $M$ distinct components the
linear system \eqref{eq:linSystemSchur} can be partitioned w.r.t. the $M$
domains
\begin{equation}
  \label{eq:partitionS}
  \widehat{S} =
  \begin{bmatrix}
    \widehat{S}_{11} & \dots & \widehat{S}_{1m} \\
    \vdots & \ddots & \vdots \\
    \widehat{S}_{1m}^\top & \dots & \widehat{S}_{mm}
  \end{bmatrix} \,, \quad m = 1,\dots,M \;.
\end{equation}
The same holds for the preconditioner. The partitioned preconditioner is
$P^{-1}_S = \diag( P^{-1}_{S_{11}}, \ldots , P^{-1}_{S_{mm}} )$. Hence, if one
component $k$ changes its position, only the off-diagonal blocks
$\widehat{S}_{km}$, $m=1,\dots,M$ with $m \neq k$ need to be updated while the
preconditioner remains unchanged.

This simplicity in handling moving parts is a preeminent advantage for the
presented FEM-BEM coupling. The application engineer is provided with maximal
convenience without loss of accuracy during the modeling process.

\section{Conclusion\label{sec:conclusion}}
This work presents an implementation of a FEM-BEM coupling scheme for
electromagnetic field simulations. The coupling eliminates problems that are
inherent to a pure FEM approach. In detail, the benefits of the FEM-BEM scheme
are: The decay conditions are fulfilled exactly, no meshing of parts of the
exterior air region is necessary, and -- most importantly -- the handling of
moving parts is incorporated in an intriguingly simple fashion. As a downside,
the coupling is considerably more complex than a pure Finite Element
scheme: The use of Boundary Element Methods requires advanced compression
techniques and the incorporation of non-simply connected domains demands
further adoptions of the scheme. However, once these issues have been handled,
the FEM-BEM formulation in conjunction with a state-of-the-art preconditioner
demonstrates its potency. The numerical tests not only reveal an accurate
convergence behavior but also prove the algorithm to be suitable for
industrial applications -- especially, if improvements such as the
incorporation of periodicities and the handling of multiple domains are
implemented.

Eventually, the presented FEM-BEM formulation is an expedient supplement to
a pure FEM. While not intended to be used under all circumstances
it represents a powerful tool in case that high accuracies together with
simple mesh-handling facilities are required. 

\bibliographystyle{amsplain}

\begin{thebibliography}{10}

\bibitem{allgower:98}
E.~L. Allgower, K.~Georg, R.~Miranda, and J.~Tausch, \emph{{N}umerical
  {E}xploitation of {E}quivariance}, ZAMM - {J}ournal of {A}pplied
  {M}athematics and {M}echanics \textbf{78} (1998), no.~12, 795--806.

\bibitem{bergot2013high}
M.~Bergot and M.~Durufl{\'e}, \emph{High-order optimal edge elements for
  pyramids, prisms and hexahedra}, Journal of Computational Physics
  \textbf{232} (2013), no.~1, 189--213.

\bibitem{coltonKress2012inverse}
D.~Colton and R.~Kress, \emph{Inverse acoustic and electromagnetic scattering
  theory}, vol.~93, Springer Science \& Business Media, 2012.

\bibitem{greengardRokhlin:97}
L.~Greengard and V.~Rokhlin, \emph{A new version of the {F}ast {M}ultipole
  {M}ethod for the {L}aplace equation in three dimensions}, Acta Numerica
  \textbf{6} (1997), 229--269.

\bibitem{hackbusch:09}
W.~Hackbusch, \emph{Hierarchische matrizen: Algorithmen und analysis},
  Springer-Verlag Berlin Heidelberg, 2009.

\bibitem{hiptmair:02FEM}
R.~Hiptmair, \emph{Finite elements in computational electromagnetism}, Acta
  Numerica \textbf{11} (2002), 237--339.

\bibitem{hiptmair:02}
R.~Hiptmair, \emph{Symmetric coupling for eddy current problems}, SIAM Journal
  on Numerical Analysis \textbf{40} (2002), no.~1, 41--65.

\bibitem{hiptmair2002generators}
R.~Hiptmair and J.~Ostrowski, \emph{Generators of
  ${H}_1({\Gamma}_h,\mathbb{Z})$ for {T}riangulated {S}urfaces: {C}onstruction
  and {C}lassification}, SIAM Journal on Computing \textbf{31} (2002), no.~5,
  1405--1423.

\bibitem{hiptmairXu:07}
R.~Hiptmair and J.~Xu, \emph{Nodal auxiliary space preconditiong in {H}(curl)
  and {H}(div) spaces}, SIAM Journal on Numerical Analysis \textbf{45} (2007),
  2483--2509.

\bibitem{hsiaoWendland:08}
G.C. Hsiao and W.L. Wendland, \emph{Boundary integral equations}, Applied
  mathematical sciences, Springer, 2008.

\bibitem{jackson99}
J.~D. Jackson, \emph{Classical electrodynamics}, 3rd ed., John Wiley \& Sons,
  Inc., 1999.

\bibitem{kurzPeriodicity:03}
S.~Kurz, O.~Rain, and S.~Rjasanow, \emph{{A}pplication of the adaptive cross
  approximation technique for the coupled {BE}-{FE} solution of symmetric
  electromagnetic problems}, Computational mechanics \textbf{32} (2003),
  no.~4-6, 423--429.

\bibitem{monk2003finite}
P.~Monk, \emph{Finite element methods for maxwell's equations}, Oxford
  University Press, 2003.

\bibitem{nedelec:80}
J.C. N{\'e}d{\'e}lec, \emph{Mixed finite elements in $\mathbb{R}^3$},
  Numerische Mathematik \textbf{35} (1980), 315--341.

\bibitem{of:06}
G.~Of, \emph{{BETI}-{G}ebietszerlegungsmethoden mit schnellen
  {R}andelementverfahren und {A}nwendungen}, Ph.D. thesis, Institut f{\"u}r
  Angewandte Analysis und Numerische Simulation, Universit{\"a}t Stuttgart,
  2006.

\bibitem{minres1975}
C.~C. Paige and M.~A. Saunders, \emph{Solution of sparse indefinite systems of
  linear equations}, SIAM Journal on Numerical Analysis \textbf{12} (1975),
  no.~4, 617--629.

\bibitem{raviart:1977}
P.~A. Raviart and J.~M. Thomas, \emph{A mixed finite element method for 2-nd
  order elliptic problems}, Mathematical aspects of finite element methods,
  Springer, 1977, pp.~292--315.

\bibitem{xu1992iterative}
J.~Xu, \emph{Iterative methods by space decomposition and subspace correction},
  SIAM review \textbf{34} (1992), no.~4, 581--613.

\end{thebibliography}

\providecommand{\bysame}{\leavevmode\hbox to3em{\hrulefill}\thinspace}
\providecommand{\MR}{\relax\ifhmode\unskip\space\fi MR }
\providecommand{\MRhref}[2]{%
  \href{http://www.ams.org/mathscinet-getitem?mr=#1}{#2}
}
\providecommand{\href}[2]{#2}

\end{document}